# ON THE NATURE OF FINANCIAL LEVERAGE

Yaroslav Ivanenko[1][2]

## 1. Introduction

Traditionally, since von Neuman [1] and Savage [2], decision theory provided the language to many topics in economics and finance [3]. Nevertheless, this link is considered by many as self-evident, and perhaps that is why the finance part of the application of decision theory stays unjustly in shadow among financial engineers. It seems that some usual economic and corporate finance notions such as Return on Capital, Leverage, Return on Investment and Price [4, 5, 6], once translated into the language of decision theory, become much clearer from pedagogical point of view. Therefore, the goal of this article is rethinking of some financial terms in terms of that post-Savage decision theory, that deals with uncertainty attitude and sets of finitely additive probabilities [7, 8, 9, 10]. Moreover, due to the direct link that exists between this decision theory and the world of statistically unstable (*nonstochastic*) random phenomena [9], this rethinking may open new possibilities for the corporate and financial risk management.

## 2. Leverage as Decision

Financial leverage is one of the essential instruments of financial activity [4]. The next example of Return on Capital calculation is taken from [6] (paragraph "Leverage: hero or zero?).

Let *C* – capital, *B*– borrowed funds, *A= C+ B* – assets, *ROI* – return on investment, *COF* – cost of funding (of *B*), *COC*- cost of capital (of *C*). Then Return on Capital (*ROC*) is

$$(1) \quad ROC = \frac{(C+B)(1+ROI) - B(1+COF) - C(1+COC)}{C}$$

For example, *C=10$, B=90$, ROI=6%, COF=COC=COST=5%*. Let *LEV = A/C=(C+B)/C = 10* denotes the leverage. Then, in this particular case,

$$(2) \quad ROC = LEV(ROI - COST)$$

In the numerical example above, *ROC=10%*. However, if *ROI=4%*, then *ROC=-10%*!

Now, let us interpret this using the language of decision systems. Return on Investment, *ROI (%)*, is, obviously, the unknown (uncontrolled, random) variable $\theta \in \Theta$, where $\Theta$ is the set of its values. The prices *COC=COF=COST* and the leverage *LEV* are the elements of decision.[3] Namely, denote *LEV=u* and *COST=p,* and $d = (u,p) \in D$, where the set of decisions $D = \mathbb{R}^+ \times \mathbb{R}^+$, with leverage $u \in U = \mathbb{R}^+$ and price $p \in P = \mathbb{R}^+$. Return on Capital (2), $ROC$ – is the consequence of the decision $d$, $L: \Theta \times D \to \mathbb{R}$, so that (2) becomes

(3) $$L(\theta, d) = u(\theta - p), \quad d = (u, p) \in D, \quad \theta \in \Theta.$$

---

[1] **Disclaimer: the ideas expressed and the results obtained in this article reflect the views of the author and are not necessarily shared by the author's institution of affiliation.**

[2] Risk management, Banque de France, 31, Rue Croix des Petits Champs, 75001 Paris, France

[3] How to include price in the decision-theoretical construct was shown as well in [11].

Expression (3) is multiplicative in leverage $u$: for the same $\theta$, one could have quite different returns for different decisions $u$. In other words, the leverage $u$ is the sensibility of the Return on Capital to the value of the Return on Investment, which is random by nature: the higher the leverage, the better we are off if things turn good, but the worse we are off if things turn bad.

We have thus a so called *matrix decision scheme* [9]

(4) $$Z = (D, \Theta, L),$$

where $D$ - is the set of decisions, $\Theta$ - is the set of values of the uncontrolled variable (state of Nature), the set of values of the function $L: \Theta \times D \to \mathbb{R}$ is the set of consequences (in our case, it coincides with the set of values of the $ROC$).

The preference relation of decision maker over the set of consequences, which in our case coincides with the set of values of the Return on Capital (2), is given by the values of this return or, by the same token, the values of the function $L(\theta, d)$ (3): the higher the value of the return on capital $L(\theta, d)$, the better.

According to decision theory [8], a decision maker estimates her decisions following the indications of the criterion

(5) $$L_Z^*: D \to \mathbb{R}.$$

In case of the expected utility theory [1, 2, 5], this criterion has the form

(6) $$L_Z^*(d) = \int v(L(\theta, d)) q(d\theta),$$

where $q(\cdot)$ is a probability distribution on $\Theta$, and $v(\cdot)$ is the utility function on consequences. In order to analyze the behavior of the investor facing a problem of choice of a decision, classical expected utility theory offers a wide spectrum of utility functions on consequences: concave for risk averse, convex for risk prone and linear for risk neutral decision makers. Hence, within this theory the shape of the utility function on consequences is crucial for the decision making. However, practitioners within financial industry do not know their utility functions and what is of interest for them – is, strictly speaking, only the consequence of their decision, the value of the return on capital. That is why, more often than not, their major interest is probabilities of consequences and, hence, statistical data. Nevertheless their behavior in one and the same situation of choice is different, even when statistical data is the same. Post-Savage decision theory offers an interesting way to bypass the issue of the unknown utility function.

If decision maker belongs to the class of uncertainty averse decision makers [7, 8, 9, 10] she, in order to evaluate her decision $d$, would use the criterion

(7) $$L_Z^*(d) = \min_{q \in Q} \int L(\theta, d) q(d\theta),$$

where $Q$ is a *statistical regularity* on $\Theta$ in the form of a closed in *-weak topology, *not necessarily convex*, family of *finitely–additive* probability distributions on $\Theta$, describing a so called *nonstochastic random* behavior of $\theta \in \Theta$ [9].[4] Taking into account (3), one obtains

(8) $$L_Z^*(d) = \min_{q \in Q} \int L(\theta, d) q(d\theta) = u \left( \min_{q \in Q} E_q(\theta) - p \right).$$

Minimum is a concave function on decisions and thus, leaving the utility on consequences coincide with the consequences themselves, this construction assimilates the risk averse behavior, which in the expected utility case was modeled by a concave utility function on consequences.

If uncertainty about the behavior of the uncontrolled parameter $\theta$ is complete, that is when $Q$ is the set of all finitely-additive probability measures on $\Theta$, including all Dirac delta distributions, then (8) degenerates in

(9) $$L_Z^*(d) = u \left( \min_{\theta \in \Theta} \theta - p \right).$$

If the investor is *uncertainty prone*, then in (8) she would use *max* rather than *min* over $Q$ when estimating her decisions [8, 10].

If it happens that minimal expected value in (8) is lower than the available for negotiation prices and, at the same time, maximal expected value is higher than the prices, then behavior of the two types of decision makers would be completely different: tending to maximize their correspondent expected profits (minimal for the uncertainty avers and maximal for the prone), the uncertainty avers would choose as low leverage as possible, solving

(10) $$\max_{d \in D} L_Z^*(d) = \max_{u \in U} u \cdot \left( \min_{q \in Q} E_q(\theta) - p \right),$$

while the uncertainty prone would choose it as high as possible, solving

(10′) $$\max_{d \in D} L_Z^*(d) = \max_{u \in U} u \cdot \left( \max_{q \in Q} E_q(\theta) - p \right).$$

If we assume that $\theta$ itself, that is the *ROI*, depends on the Return on Capital of a leveraged vehicle, then *ROC* (2)-(3) becomes dependent on the previous leverage. Indeed, setting for convenience $\theta_i = \theta_i - p$, and admitting that the dependence is linear,

(11) $$\theta \sim L(\theta_1, d_1) = u_1 \theta_1,$$

then, provided such leverage is done consecutively *N* times,

(12) $$\theta_i \sim L(\theta_{i+1}, d_{i+1}) = u_{i+1} \theta_{i+1}, i = 0, \ldots, N - 1,$$

one obtains

---

[4] Unlike [7] and [8], where $Q$ is a *convex* closed set of finitely-additive probabilities, interpreted as a set of prior distributions, in this paper, due to the theorem of existence of statistical regularities of nonstochastic (statistically unstable) random phenomena [9], we adhere to the interpretation of such closed, not necessarily convex, set as the statistical law of actual random behavior of the *ROI*.

$$\text{(13)} \qquad L(\theta, d) = u\theta \sim uu_1\theta_1 \sim \ldots \sim \prod_{i=0}^{N} u_i\, \theta_N \;,$$

where $u_i$ – is the leverage of the $i$th Return on Capital. Hence, the criterion (8) becomes

$$\text{(14)} \qquad L_Z^*(d) = \min_{q \in Q} \int L(\theta, d) q(d\theta) \sim [\prod_{i=0}^{N} u_i] \min_{q \in Q_N} E_q(\theta_N),$$

where $Q_N$ is now the statistical regularity of the primitive random variable $\theta_N \in \Theta_N$.

The multiplicative nature of leverage seems to be the object of the analysis of structured finance vehicles proposed in [6], where the product $\prod_{i=0}^{N} u_i$ seems to be given the name of the *see-through leverage, and* the losses in the subprime loans portfolios was the primitive random variable. As it were, the sensitivity to this variable in some structures reached 1000s, witnessing of extremely high sensitivity to randomness within the financial system.

### 3. Conclusions

We see that the language of decision systems allows for an effective formalization of about any financial decision making situation. In particular, we rapidly detect the multiplicative nature of financial leverage and the resulting amplification of risk. The notion of statistical regularity of nonstochastic randomness allows, in particular, abandoning utility functions on consequences as analytic tool and permits working directly with profits and losses, taking at the same tame into account such an important behavioral feature as *uncertainty attitude* [7, 8, 9, 10]. At the same time, the notion of statistical regularities of nonstochastic randomness [9] allows for an alternative look on leverage as not so much "a sensitivity to model assumptions" [6], but more likely as a *sensitivity to randomness*: risk averse decision makers lower this sensitivity, risk prone make it higher. This decision-theoretical classification corresponds to the simple observation that there are people who prefer living within their means and people who prefer living on credit. Together with the knowledge of statistical regularity of the random phenomenon at hand, the recognition by the decision maker of her rationality class is crucial for practical decision making.

### Acknowledgements

The author thanks Akaki Gabelaia from Tbilisi State University of Economic Relations, Georgia, and to Bertrand Munier, Sorbonne's Business School, Paris, for very useful discussions and support.